\documentstyle[12pt]{article}
\title{}
\author{}
\begin{document}
\large

\baselineskip=8mm plus 1mm minus 1mm

\begin{center}
{\bf PREONS, DARK MATTER AND THE PRODUCTION OF EARLY COSMOLOGICAL STRUCTURES}

\vspace*{.5cm}

V.Burdyuzha${}^{1}$, O.Lalakulich${}^{2}$, Yu.Ponomarev${}^{1}$,
G.Vereshkov${}^{2}$

\vspace*{.5cm}
{\it
${}^{1}$ Astro Space Center of Lebedev Physical\\
Institute of Russian Academy of Sciences\\
Profsoyuznaya 84/32, 117810 Moscow, Russia\\
${}^{2}$ Rostov State University,\\
Stachki str. 194, 344104 Rostov on Don, Russia\\
}

\end{center}

\vspace*{1cm}
\abstract{
If the preon structure of quarks, leptons and gauge bosons will be
proved then in the Universe during relativistic phase transition the
production of nonperturbative preon condensates has been occured collective
excitations of which are perceived as pseudogoldstone bosons. Dark matter
consisting of pseudogoldstone bosons of familon type contains a "hot"
component from massless particles and a "cold" (nonrelativistic) component
from massive particles. It is shown that such dark matter was undergone
to two relativistic phase transitions temperatures of which were different.
In the result of these phase transitions the structurization of dark matter
and therefore the baryon subsystem has taken place. Besides, the role of
particle generations in the Universe become more evident. For the possibility
of structurization of matter as minimum three generations of particles are
necessary.}

\newpage

\noindent
{\bf 1. INTRODUCTION}

Observational data of distant objects (quasars at $z > 4$ [1]
and CO molecular  lines
at  $z \approx 4.69$ [2]) show that some baryon objects and baryon
large-scale structure (LSS) were created at least at redshifts $z
\sim 6 \div 8$. This is the difficulty for standard CDM and CHDM
models to produce their and in more early epoches $\Lambda$CDM model
describes LSS formation better.  The conventional scenario to explain
the formation of the most structures at $z \sim 2 \div 3$ is the
bottom-up hierarchical formation scenario such as CDM [3]. If early baryon
cosmological structures were produced on z more than 10 then the key role
in this process must play dark matter (particles forming more than 90 \%
of the Universe mass).

In the standard model dark matter (DM) consists of ideal gas of light
$m \approx 0$ particles practically noninteracting with usual matter (till
now they are not detected  because of their superweak interaction with
baryons and leptons[4]). In a new theory of DM an effect of strong
structurization of DM must be inevitably in early epoches.
The problem is to find the character of this phenomena.

In the frame of purely phenomenological approximation one can propose that
values of initial fluctuations in DM were not connected with the value of
relic radiation fluctuations. In this case the selection of the
values of DM initial fluctuations can provide any moment of structure
production.  Authors of articles [5-6] have
considered DM as the gas of pseudogoldstone bosons (PGB) in which a
relativistic phase transition (RPT) is possible at superlow (comparing
with the usual scale of elementary particle physics)
temperature. The fact of this phase transition proposes strong nonlinearity
of DM, that is these PGB must strongly interact with each other. In the
mirror models this property is  realized automatically. In the gas of PGB
particles this phenomenon is exotic and it imposes strict limits on the
hypothesis about the nature of PGB.

PGB as physical objects arise as the result of the spontaneous  breaking of
the continuous symmetries of vacuum. In modern theories of elementary
particles four types of PGB are discussed: axions, arions, familons, and
majorons. The small masses of PGB arise in the result of superweak interactions of
Goldstone fields with nonperturbative vacuum condensates.
The values of these mass are limited by astrophysical
data [7]:
$$m_{PGB} \sim 10^{-3} \div 10^{-5} \;\; eV \eqno(1.1)$$
(laboratory experiments admit masses of PGB  to $\sim 10 \; eV$ [7]).
The estimation (1.1) has been got from assumption that PGB are elements
of DM influencing on dynamics of the Universe expansion and on the
process of production of baryon LSS.

The emergence of massive terms in Lagrangian of Goldstone fields
corresponds formally mathematicaly the effect of the production of PGB
masses. From general considerations it is possible to propose, that,
dependending on the PGB type and the structure of nonperturbative
vacuum, massive terms can arise as with "right" as with "wrong" sign. This
sign predetermines the destiny of residual symmetry of pseudogoldstone
fields. In the case of wrong sign for low temperatures
$$
T < T_{c} \sim m_{PGB} \sim (0.1 \div 10^{5}) \; K \eqno(1.2)
$$
a Goldstone condensate produced inevitably and the symmetry of the
vacuum state breaks spontaneously. The appearence of a
condensate at $T = T_{c}$ and more low temperatures is a relativistic phase
transition (RPT) from the high symmetric (HS) phase in low symmetric (LS)
phase of PGB gas. As it is known, the general theory of cosmological RPT was
formulated by Kirzhnits and Linde [8].

The idea of RPT in the cosmological gas of PGB in the connection with baryon
large scale structure problem was formulated by Frieman et al. [6]. In this
article we discuss the properties of cosmological gas containing PGB
of familon type and investigate the preon-familon model of this RPT
quantitatively. The familon symmetry is experimentally observed as the fact,
that different generations of quarks and leptons absolutely identically
participate in all gauge interactions. The breaking of familon symmetry is
manifested in the values of particle masses of different generations.

There were proposed some models (see as the example [9]) where the familon
symmetry is the horisontal gauge symmetry,  breaking down by Higgs
condensates. The hypothesis is quite natural because it is based on
general ideas about the unification of particles and interactions.

However for  existence of familons as physical objects it is necessary
that at least some Goldstone degrees of freedom were not transfered to
vector states as their longitudinal polarizations.  If
this condition is completely fulfilled then familon fields are complex
and they possess residual global  $U{(1)}$ symmetries. If these familon
fields have mass terms with wrong sign then these residual
symmetries must be spontaneously broken as temperature has decreasd.
The properties of any pseudogoldstone bosons (as pseudogoldstone
bosons of familon type) depends on physical realization of Goldstone
modes. These modes can be arisen from fundamental Higgs fields
(as it is done in the work [9])
or from collective excitations of a heterogenic nonperturbative vacuum
condensate, more complicated than  quark-gluon one in QCD. The second
possibility can be realized in the theory in which quarks, leptons and
intermediated bosons are composite objects.
Such model is called the preon model of elementary particles. Thus we will discuss
the properties of familon gas predicted by the preon theory, and will
investigate the cosmological consequences of the simplest boson-familon
model [10]. Our interest to the preon model is induced by the
possible  interpretation of HERA experiment [11] as
the leptoquark resonance if a leptoquark is the composite object.

RPT in the cosmological familon gas is a phase transition essentially of the
first order with a wide temperature region of HS and LS phase coexistence.
The numerical modelling of this RPT has shown that the space
interchange of HS and  LS phases with the contrast of the energy
density $\delta \rho/ \rho \sim 1$ arises in the Universe in the epoch
of the phases coexitence. The characteristic scale of the block-phase
structure is defined by the distance to horizon in RPT moment, that is
the large-scale structure of DM is produced. The baryon subsystem
duplicates the DM large-scale structure due to gravitational interaction
with DM.
To explain the hierarchy of modern stuctures our model describes
at least two RPT at temperatures no more than a few electronvolt,
one phase transition occures at the postrecombination epoch.

\vspace*{.3cm}

\noindent
{ \bf 2. THE PHYSICAL NATURE OF FAMILONS AS THE EXCITATIONS OF
NONPERTURBATIVE VACUUM}

As we have already noted, the breaking of familon symmetry makes itself
evident in the spliting of quarks and leptons masses. Therefore the
discussion of the physical nature of familons must appeal to the
problem of the origin of fermion masses. There are two
mechanisms of mass generation: Higgs and nonperturbative one, in which
the masses generation are the result of fermion fields interactions with
non-perturbative vacuum condensates (in the simplest case they are
quadratic on quantum fields).
In QCD the second mechanism is drawn for the mass
generation of $u$ and $d$ quarks. The effect of mass generation of
quarks is illustrated by the diagram:
\\
\begin{picture}(350,110)
\put(100,50){\vector(1,0){15}}
\put(115,50){\line(1,0){30}}
\put(200,50){\vector(1,0){30}}
\put(230,50){\line(1,0){15}}
\put(145,50){\line(0,-1){30}}
\put(200,50){\line(0,-1){30}}
\put(142,20) {$ \mbox{x} $}
\put(142,75) {$ \mbox{x} $}
\put(197,20) {$ \mbox{x} $}
\put(197,75) {$ \mbox{x} $}
\put(380,50) {$ \mbox{(2.1)} $}
\multiput(145,50)(0,3){8}{\oval(3,6)}
\multiput(200,50)(0,3){8}{\oval(3,6)}
\put(130,90){$\langle 0 | \frac{\alpha_{s}}{\pi}
G^{a}_{\mu \nu} G^{\mu \nu}_{a} | 0 \rangle $}
\put(130,5){$\langle 0 | \bar{q}_{L} q_{R} +
\bar{q}_{R} q_{L} | 0 \rangle $}
\put(95,55) {$q_L $}
\put(220,55) {$q_R $}
\put(280,50) {$q=u,\ d $}
\end{picture}
\vspace*{.5cm}

The numerical values of gluon and quarks condensates standing in (2.1)
are known from experimental data [12]:
$$
\begin{array}{c}
\langle 0 \mid \frac{\alpha_{s}}{\pi} G^{a}_{\mu \nu} G^{\mu \nu}_{a} \mid
0 \rangle = (360 \pm 20 \; MeV)^{4} \\
\hspace*{13cm} (2.2) \\
\\
\langle 0 \mid \bar{u} u \mid 0 \rangle = \langle 0 \mid \bar{d} d
\mid 0 \rangle = - (225 \pm 25 \; MeV)^{3}
\end{array}
$$
The basic characteristic of QCD vacuum is the gluon condensate,
originating from the
nonperturbative fluctuations of gluon fields being arisen in the
processes of gluon vacuum tunelling among degenerated on energy states
with different topological numbers [12]. The quark condensates are induced
by the fluctuations of initially massless quark fields appearing as a reaction
of quark fields on vacuum fluctuations of gluon fields. The initial massless
of $u, d$ quarks provides the essentially high intensity of these reactions
that represents the value of the quark condensate is smaller but not very
small comparing with the value of gluon condensate.

In the frame of QCD the masses of heavy quarks ($s, c, b, t$) have not
the analogous interpretation. Really, QCD predicts that magnitudes of
heavy quarks vacuum condensates are in inverse proportion to the
meaning of their mass:
$$\langle 0 \mid \bar{Q} Q \mid 0 \rangle \approx - \frac{1}{12 m_{Q}}
\; \langle 0 \mid \frac{\alpha_{s}}{\pi} G^{a}_{\mu \nu} G^{\mu
\nu}_{a} \mid 0 \rangle, \;\;\; Q = s, c, b, t. \eqno(2.3)$$

But in the nonperturbative mechanism masses of quarks are proportional
to the value of the corresponding condensate. In this case it is
necessary to attract the additional physical considerations for the
explanation of
spectra of heavy quark masses. Phenomenologically these masses can be
brought using Higgs mechanism, althougt the additional introduction of a
great number of constants of unknown nature is requested. But Higgs bosons
are not detected till  now while signals a new physics have created
[11].

Let us consider now the simplest boson-fermion  preon
model of left-handed quarks and leptons. The basic elements of this
model are:
chiral fermionic preons of up $U^{\alpha}_{L}$ and down $D^{\alpha}_{L}$
type, and scalar
preons of quarks $\phi^{i \alpha}_{a}$ and leptons $\chi^{i
\alpha}_{e}$ types. Here $i$ is the color index of QCD; $a,b,c = 1,
2, 3, \; l, m = 1, 2, 3$ are numbers of quark and lepton generations;
$\alpha$ is the index of metacolor corresponding to a new
metachromodynamical interaction, connecting preons in quarks and
leptons. The following is the structural formula illustrating the interior
structure of elementary particles:
$$
\begin{array}{l c l c l c l}
u^i_{La} & = & U^{\alpha}_{L} \phi_{a}^{+i\alpha} & \hspace{1.0 cm} &
u_{La}^{i} & = & (u_L^i,\ c_L^i,\ t_L^i) \\
d^i_{La} & = & D^{\alpha}_{L} \phi_{a}^{+i\alpha} & \hspace{1.0 cm} &
d_{La}^{i} & = & (d_L^i,\ s_L^i,\ b_L^i) \\
\nu^i_{Ll} & = & U^{\alpha}_{L} \chi_{l}^{\alpha} & \hspace{1.0 cm} &
\nu_{Ll}^{i} & = & (\nu_{Le},\ \nu_{L\mu},\ \nu_{L\tau}) \\
l^i_{Ll} & = & D^{\alpha}_{L} \chi_{l}^{\alpha} & \hspace{1.0 cm} &
l_{Ll}^{i} & = & (e_{L},\ \mu_{L},\ \tau_{L}) \\
\end{array}
\eqno(2.4)
$$
$$(LQ)_{al} = \phi^{+ia}_{a} \chi^{\alpha}_{l} \eqno(2.5)$$

The structure of the good known particles  is the formulae (2.4). The
structure of leptoquarks is the formula (2.5).

The interior of quarks and leptons the metagluon fields $G^{\omega}_{\mu
\nu}$ and scalar preon fields are the confinement state. This effect
on itself physical nature is identical with confinement of quarks and
gluons interior of hadrons and  it is provided by the existence of
nonperturbative metagluon and preon condensates
$$ \langle 0 \mid \frac{\alpha_{mc}}{\pi} \; G^{\nu}_{\mu \nu} G^{\mu
\nu}_{\nu} \mid 0 \rangle \sim \Lambda^{4}_{mc} \eqno(2.6)$$
$$\langle 0 \mid \phi^{+ia}_{a} \phi^{ia}_{b} \mid 0 \rangle = V_{ab}
\sim - \Lambda^{2}_{mc} \eqno(2.7)$$
$$\langle 0 \mid \chi^{+a}_{l} \chi^{a}_{m} \mid 0 \rangle = V_{lm}
\sim - \Lambda^{2}_{mc}.\eqno(2.8)$$

The condensates (2.6) and (2.7) in combination with gluon and quark
condensates (2.2) and (2.3) provide the  mechanism of mass production
of quarks of all three generations

\begin{picture}(350,200)
\put(50,100){\vector(1,0){10}}
\put(60,100){\line(1,0){10}}
\put(70,100){\circle*{6}}
\put(70,103){\line(1,0){35}}
\multiput(70,97)(10,0){4}{\line(1,0){5}}
\multiput(105,93)(0,-10){7}{\line(0,1){5}}
\multiput(200,150)(0,-10){13}{\line(0,1){5}}
\put(105,150){\vector(0,-1){25}}
\put(250,150){\vector(0,-1){25}}
\put(105,125){\line(0,-1){22}}
\put(250,125){\line(0,-1){25}}
\put(250,100){\vector(1,0){40}}
\put(290,100){\line(1,0){15}}
\multiput(108,115)(3,0){11}{\oval(6,3)}
\multiput(108,70)(3,0){11}{\oval(6,3)}
\multiput(150,100)(3,0){17}{\oval(6,3)}
\put(140,112) {$ \mbox{x} $}
\put(140,97) {$ \mbox{x} $}
\put(140,67) {$ \mbox{x} $}
\put(103,150) {$ \mbox{x} $}
\put(103,25) {$ \mbox{x} $}
\put(247,150) {$ \mbox{x} $}
\put(197,150) {$ \mbox{x} $}
\put(197,25) {$ \mbox{x} $}
\put(365,100) {$ \mbox{(2.9)} $}
\multiput(80,40)(3,0){8}{\oval(6,3)}
\multiput(203,40)(3,0){22}{\oval(6,3)}
\multiput(80,10)(3,0){62}{\oval(6,3)}
\multiput(80,10)(0,3){10}{\oval(3,6)}
\multiput(270,97)(0,-3){10}{\oval(3,6)}
\put(267,37) {$ \mbox{x} $}
\put(267,60) {$ \mbox{x} $}
\put(267,7) {$ \mbox{x} $}
\put(100,175) {$\langle 0 | \bar{U}^{\alpha}_{L}\phi^{r\alpha}_{c} u^r_{Rb} | 0 \rangle
\equiv
\langle 0 | \bar{u}^m_{Lc} u^m_{Rb} | 0 \rangle $}
\put(110,17) {$\langle 0 | \phi^{+k\beta}_{a} \phi^{k\beta}_{c} | 0 \rangle $}
\put(110,80) {$\langle 0 | \frac{\alpha_{mc}}{\pi}
G^{\alpha \gamma}_{\mu \rho} G^{\gamma \beta}_{\rho \nu}
G^{\beta \alpha }_{\nu \mu} | 0 \rangle $}
\put(255,50) {$\langle 0 | \frac{\alpha_{s}}{\pi}
G^{ik}_{\mu \nu} G^{kr}_{\nu \rho}
G^{ri}_{\rho \mu} | 0 \rangle $}
\put(50,110) {$\bar{u}^{i}_{L\alpha} $}
\put(280,110) {$u^{i}_{Rb} $}
\put(80,110) {$\bar{U}^{\gamma}_{L} $}
\put(80,80) {$\phi^{i\gamma}_{\alpha} $}
\end{picture}
\vspace*{.3cm}

\noindent
here $G^{ik}_{\mu \nu} = \lambda^{ik}_{n} G^{in}_{\mu \nu}, \;\;
\lambda^{in}_{n}$ are Gell-Mann matrixes; $G^{\alpha \beta}_{\mu \nu} =
\lambda^{\alpha \beta}_{\omega} G^{\omega}_{\mu \nu}, \;
\lambda^{\alpha \beta}_{\omega}$ are the analog of a Gell-Mann matrix for
metacolour.

As it is seen from (2.9) the basic contribution in the breaking of
vacuum familon symmetry is formed by preon condensates (2.7).
For a realistic vacuum state the condensate matrixes must be brought to
diagonal form. From the suggestion (2.1) the equality $V_{11} = 0$
follows. Thus on the preon level the chiral-familon symmetry $SU(2)_{L}
\times SU(2)_{R}$ joining quarks of the second and  third generations
must be considered.

The importantest prediction of diagram (2.9) is that  the interactions
of quark excitations with complex condensates containing preon and
quark-preon components must be considered for full discription of the
production mass effect
$$
M^{(u)}_{ab} = \langle 0 \mid \phi_a^{\alpha k}\phi_c^{\alpha k}
\bar{U}_L^{\beta} \phi_c^{\beta i} q_{Rb}^{i} \mid 0 \rangle
\eqno(2.10)
$$
$$
M^{(d)}_{ab} = \langle 0 \mid \phi_a^{\alpha k}\phi_c^{\alpha k}
\bar{D}_L^{\beta} \phi_c^{\beta i} q_{Rb}^{i} \mid 0 \rangle
\eqno(2.11)
$$
In (2.10) and (2.11) for simplicity the gluon and metagluon condensates
as sources of nonperturbative preon condensates are not brought. Their
role is evident from diagram (2.9). Therefore the theory of preon
predicts a complicated structure of the heterogenic nonperturbative
vacuum. The element (2.10) from this structure participates in the
process of mass production of up quarks from the second and third
generations, for down quarks the element (2.11) does the same. The
analogous analysis of generation mass process for charge leptons of the
second and third generations gives yet one element of the interior
structure of vacuum
$$
M^{(l)}_{lm} = \langle 0 \mid \chi^{\alpha}_{l} \chi^{\alpha}_{r}
\bar{D}^{\beta}_{L} \chi^{\beta}_{r} l_{Rm} \mid 0 \rangle
\eqno(2.12)
$$
Formally formulae (2.10), (2.11) and (2.12) are  2 x 2 matrixes of
general form containing 8 numbers. It is easy to calculate that a
number of parameters in real and diagonal matrixes (2.10), (2.11)  and
(2.12) equals a number of particles containing in the second and third
quark and lepton generations.

Familons as physical objects are collective excitations of nonperturbative
condensates (2.10), (2.11) and (2.12). These excitations are the result
of local processes of weaking and rebuilding of correlations among fields.
Three types of condensates correspond to three families of familon fields.
Evidently, also that the number of familons in a family equals eight
that is it coinsides with a maximum numbers of parameters of condensate
matrixes. In each family two familon fields are arisen as the consequence
of the  local perturbation of condensate  density energy. If a familon
family is described by complex 2 x 2 matrixes of general form then
perturbation of density energy will correspond to diagonal elements of
this matrix. Perturbations of energy density have a mass around scale of
metacolour confinement $\Lambda_{mc}$.

The rest six familons of a family are arisen as a results of rebuilding
of a condensate. A small mass of rest the Goldstone modes gain
interacting with quark condensates (their status in this case are PGB).
On the diagram (2.9) the mass of quarks is formed by interaction of
quantum components of quark subsystems with condensate components of
preon subsystems. In the similar spirit a small mass of pseudogoldstone
familons is gained for interaction of perturbations of condensate preon
subsystems with condensate components of quark subsystems (2.3).

For cosmological applications of the familon theory based on the
hypothesis about preon structure of vacuum and elementary particles the
most interest presents the predictions related to DM. In frame of this
theory DM is interpretated as the system of familon collective
excitations of the heterogenic nonperturbative vacuum. This system
contains three subsystems: familons of upper-quark type; familons of
lower-quark type and familons of lepton type.  On stages of the
cosmological evolution which are far from quarkonization and
leptonization of preon plasma ($T \le \Lambda_{mc}$) heavy unstable
familons are absent as evidently. On these stages each familon
subsystems are described by five field degrees of freedom. The destiny
of each subsystems of low energetic familon gas depends radical
mean on the sign of rest mass squares which are generated by
interactions of familons with quark condensates.

\vspace*{.3cm}

\noindent
{\bf 3. THE INEVITABILITY OF RESIDUAL SYMMETRY SPONTANEOUS BREAKING OF
PSEUDOGOLDSTONE FAMILON FIELDS}

The presentation about physical nature of familons is uniquely
formalized in a theoretico-field model. For simplicity the model of
familon subsystem corresponding to up quarks of the second and the
third generations $Q = (c, t)$ is considered. Because of the chiral
nature of quarks is the good known fact then the symmetry group of
model must be only the chiral-familon group $SU(2)_{L} \times
SU(2)_{R}$. The familon excitations must be described by the eight
measure reducible representation of this group factorized on two
irreducible representations ($F, fa); (\psi, \varphi_{a}$) which differ each
other by a sign of space chirality
$$L = \frac{1}{2} (\partial_{\mu} \psi \partial^{\mu} \psi +
\partial_{\mu} \varphi_{a} \partial^{\mu} \varphi_{a} + \partial_{\mu}
F \partial^{\mu} F + \partial_{mu} f_{a} \partial^{\mu} f_{a}  +
\frac{1}{2} \mu^{2}_{1} (\psi^{2} + \varphi_{a} \varphi_{a}) +$$
$$+ \frac{1}{2} \mu^{2}_{2} (F^{2} + f_{a} f_{a}) - \frac{1}{4}
\lambda_{1} (\psi^{2} + \varphi_{a} \varphi_{a})^{2} - \frac{1}{4}
\lambda_{2} (F^{2} + f_{a} f_{a})^{2} - \frac{1}{2} \lambda_{12}
(\psi^{2} + \varphi_{a} \varphi_{a})(F^{2} + $$
$$+ f_{a} f_{a}) - \frac{1}{2} \lambda_{0} (\psi F + \varphi_{a}
f_{a})^{2} - g_{1} \bar{Q} (\psi + i \gamma_{5} \tau_{a} \varphi_{a}) Q +
g_{2} \bar{Q} (\tau_{a} f_{a} - i \gamma_{5} F) Q \eqno(3.1)$$

The  quark fields are presented in (3.1) by induce nonperturbative
fluctuations (2.3) only which describe phenomena taking place outside a
hadron space. The terms with mass parameters $\mu^{2}_{1}, \mu^{2}_{2}
\sim \Lambda^{2}_{mc}$ are included in (3.1) with "wrong" signs to
provide the spontaneous breaking of familon symmetry on the scale of
metacolor confinement. In a condensate the mean values of scalar (non
pseudoscalar) fields  drop out:
$$
\langle \psi \rangle = v, \;\; \langle f_{3} \rangle = u, \eqno(3.2)
$$
that provides the conservation of parity in strong interactions. The
investigation of model (3.1) in detail has shown that these vacuum
expectations break the initial chiral-familon symmetry to the residual $U(1)$
symmetry, corresponding to conservation of the quark flavors
$$
SU(2)_{L} \times SU(2)_{R} \rightarrow U(1) \eqno(3.3)
$$

The condensates of scalar fields (3.2) generate according to the
structure of two last terms (3.1) splitted spectrum of quark masses (on
the preon level this effect is described by (2.9))
$$m_{ab} Q_{a} Q_{b} = m_{c} \bar{c} c + m_{t} \bar{t}t$$
$$m_{c} = g_{1} v - g_{2} u, \;\;\; m_{t} = g_{1} v + g_{2} u \eqno(3.4)$$

As it is seen from (3.4) for $v, u \sim \Lambda_{mc} \gg 1 \; Tev$ the
experimentally observable values of masses $\bar{m}_{c} \approx 1.3 \;
Gev, \bar{m}_{t} \approx 175 \; Gev$ are only created for very small
values of constants $g_{1}, g_{2}$ of quark-familon interactions.

A smallness of these constants provides the hyperweakness of familon
interactions with usual matter. The preon model allows an understanding
of the reason of this smallness. Really, as it is seen from (2.10) and
(2.11) a complex heterogeneous condensate is responsible for generation of
quark masses is arisen as the result of correlations of fluctuation
fields belonged to strongly different levels of structure. The
chromodynamical fluctuations on the scale $\Lambda_{c} \sim 100 \; Mev$
must be correlated with  metachromodynamical ones on the scale
$\Lambda_{mc} \gg 1 \; Tev$. Evidently that phenomenological constants
$g_{1}, g_{2}$ is proportional the probability of these correlations.
Therefore a smallness of these constants reflects a small probability of
correlations of different scale fluctuations.

Equations for vacuum expectations fixing spontaneous breaking of
familon symmetry contain quark vacuum condensates:
$$\mu^{2}_{1} v - \lambda_{1} v^{3} - \lambda_{12} u^{2} v -
g_{1} (\langle cc \rangle + \langle ll \rangle) = 0$$
$$\mu^{2}_{2} u - \lambda_{3} u^{3} - \lambda_{12} v^{2} u + g_{2}
(\langle cc \rangle - \langle ll \rangle ) = 0 \eqno(3.5)$$

These equations are used for extraction from Lagrangian (3.1) of
quantum components of familon fields. First and foremost we have
interest to the spectrum of familon masses. The spontaneous breaking
of the symmetry (3.3) transforms the six parametric group of symmetry
in one parametric one. According to the Goldstone's theorem 5 degrees
of freedom must be practically massless on scale $\Lambda_{mc}$.
Therefore 3 heavy familons must be in this model also. Two from their
are perturbations of condensate density  energies (3.2) and they
are described by orthogonal superpositions of the quantum components of
scalar fields $\psi$ and $f_{3}$:
$$M^{2}_{\psi^{'}} \approx 2 (\lambda_{1} v^{2} + \lambda_{2} u^{2});
\;\; M^{2}_{f^{'}} \approx \frac{2(\lambda_{1} \lambda_{2} -
\lambda_{12}) u^{2} v^{2}}{\lambda_{1} v^{2} + \lambda_{2} u^{2}} \eqno(3.6)$$

The third heavy familon is identificated with one from two orthogonal
superpositions of pseudoscalar fields $F$ and $\varphi_{3}$ (the second
superposition  goes to pseudogoldstone sector) and its mass is
$$M^{2}_{F^{'}} \approx \lambda_{0} (u^{2} + v^{2})$$

The pseudogoldstone modes can be constituted as a real pseudoscalar
field with mass:
$$m^{2}_{\varphi^{'}} \approx \frac{1}{24(u^{2} + v^{2})} \; \langle
\frac{\alpha_{s}}{\pi} G^{m}_{\mu \nu} g^{\mu \nu}_{n}
\rangle \left [ \frac{(m_{t} + m_{c})^{2}}{m_{c} m_{t}} - \frac{(m_{t}
- m_{c})^{2}}{m_{c} m_{t}} \right] =$$
$$= \frac{1}{6 (u^{2} + v^{2})} \; \langle \frac{\alpha_{s}}{\pi}
G^{m}_{\mu \nu} g^{\mu \nu}_{n}
\rangle  \eqno(3.7)$$
a complex pseudoscalar field with mass:
$$m^{2}_{\varphi} = \frac{1}{24 v^{2}} \; \langle
\frac{\alpha_{s}}{\pi} G^{m}_{\mu \nu} G^{\mu \nu}_{n} \rangle
[\frac{(m_{t} + m_{c})^{2}}{m_{c} m_{t}}] \eqno(3.8)$$
and a complex scalar field whose square of a mass is negative:
$$m^{2}_{f} = - \frac{1}{24 u^{2}} \; \langle \frac{\alpha_{s}}{\pi}
G^{m}_{\mu \nu} G^{\mu \nu}_{n} \rangle [\frac{(m_{t} -
m_{c})^{2}}{m_{c} m_{t}}] \eqno(3.9)$$

The complex fields with masses (3.8) and (3.9) are nontrivial
representations of the group of residual symmetry $U(1)$, real fields
with more masses than (3.6) and the field with the small mass (3.7)
are unit representations of this group. These masses are estimated as:
$$M_{\psi^{'}}, M_{f^{'}}, M_{F^{'}} \sim \Lambda_{mc} \eqno(3.10)$$
$$m_{\varphi^{'}} \sim \frac{\Lambda^{2}_{c}}{\Lambda_{mc}} \eqno(3.11)$$
$$m_{\varphi} \mid m_{f} \mid \sim \frac{\Lambda^{2}_{c}}{\Lambda_{mc}}
\sqrt{m_{t}/m_{c}} \eqno(3.12)$$

As we noted before light pseudogoldstone bosons with masses (3.10),
(3.11), (3.12) must be included in composition of DM. The negative
square mass of complex scalar field means that for low temperatures
$$T < T_{c(u)} \sim \mid m_{f} \mid \sim
\frac{\Lambda^{2}_{c}}{\Lambda_{mc}} \sqrt{m_{t}/m_{c}} \eqno(3.13)$$
PGB vacuum is unstable. That is for the temperature $T = T_{c(u)}$ the
relativistic phase transition in the state with spontaneously breaking
reaidual symmetry $U(1)$ must take place necesserily (this is
the dynamical realization of vacuum properties).  Two other familon
subsystems are studied by the analogous method. The down quark of familon
subsystem consists of pseudoscalar familons with positive square masses
$$m^{2}_{\varphi^{'}(d)} \approx \; \frac{1}{24 (u^{2}_{d} + v^{2}_{d})}
\; \langle \frac{\alpha_{s}}{\pi} \; G^{m}_{\mu \nu} G^{\mu \nu}_{n} \rangle
[\frac{(m_{b} + m_{s})^{2}}{m_{b} m_{s}} - \frac{(m_{b} -
m_{s})^{2}}{m_{b} m_{s}} ] =$$
$$= \frac{1}{6(u^{2}_{d} + v^{2}_{d})} \; \langle
\frac{\alpha_{s}}{\pi} G^{m}_{\mu \nu} G^{\mu \nu}_{n} \rangle \eqno(3.14)$$
$$m^{2}_{\varphi(d)} = \frac{1}{24 v^{2}_{d}} \; \langle \frac{\alpha_{s}}{\pi}
G^{m}_{\mu \nu} G^{\mu \nu}_{n} \rangle \; [\frac{(m_{b} +
m_{s})^{2}}{m_{b} m_{s}}] \eqno(3.15)$$
and scalar familons square mass of which is negative
$$m^{2}_{f(d)} = - \frac{1}{24u^{2}_{2}} \; \langle \frac{\alpha_{s}}{\pi}
G^{m}_{\mu \nu} G^{\mu \nu}_{n} \rangle \; [\frac{(m_{b} -
m_{s})^{2}}{m_{b} m_{s}}] \eqno(3.16)$$

 The latter means that the downquark familon subsystem for low-temperatures
is unstable also and for $T = T_{c(d)} \sim \mid m_{f(d)} \mid$ the
relativistic phase transition in the state with spontaneously breaking
residual symmetry $U(1)$ must be realized. From foregoing results it
is seen that quantitative characteristics of familon subsystems are in
fact controlled by only one parameter - the scale of metacolour
confinement. The values of all other parameters is possible to take
from an experiment or theoretical predictions of QCD. Unfortunately
analizing the subsystem of lepton familons we meet with experimental
unknown values - lepton condensates. The values of these condensates,
however, can be parametrized by masses  muon ($m_{\mu}$), $\tau$-lepton
($m_{\tau}$) and two renorm parameters $\mu_{\mu}, \mu_{\tau}$:
$$\langle \bar{\mu} \mu \rangle = - \frac{1}{4 \pi^{2}} \; m^{3}_{\mu}
ln \; \frac{m^{2}_{\mu}}{\mu^{2}_{\mu}}, \;\;\; \langle \bar{\tau} \tau \rangle = -
\frac{1}{4 \pi^{2}} \; m^{3}_{\tau} ln \frac{m^{2}_{\tau}}{\mu^{2}_{\tau}} \eqno(3.17)$$

Using (3.17) it is easy to show that PGB sector of lepton familons
consists of pseudoscalar particles square of masses of which is positive
for  $m_{\tau} > \mu_{\tau}$
$$m^{2}_{\varphi^{'}_{lep}} \approx \frac{1}{4 \pi^{2} (u^{2}_{lep} +
v^{2}_{lep})} \; m^{3}_{\tau} m_{\mu} ln \;
\frac{m^2_{\tau}}{\mu^{2}_{\tau}} \eqno(3.18)$$
$$m^{2}_{\varphi_{lep}} \approx \frac{1}{8 \pi^{2} v^{2}_{lep}}
\; m^{4}_{\tau} ln \; \frac{m^{2}_{\tau}}{\mu^{2}_{\tau}} \eqno(3.19)$$
and scalar particles having for the same conditions negative square of masses
$$m^{2}_{f_{lep}} \approx - \frac{1}{8 \pi^{2} u^{2}_{lep}} \;
m^{4}_{\tau} ln \; \frac{m^{2}_{\tau}}{\mu^{2}_{\tau}} \eqno(3.20)$$

Thus from mass formulae it is seen that DM consisting of familon type
PGB is a manycomponent heterogenetic system evolving by complicated
thermodynamical way.
In the content of DM particles with 9 different masses of rest are included.
During evolutiion this system undergone to two relativistic phase
transitions temperatures of which can differ each other more. Note that
after RPT each from complex fields decay on two real fields one from which is
massless and the second has a mass $\sim \mid m_{f(u),(d)} \mid$. The
each complex pseudoscalar field decay also on two real with masses
which are different from zero. Thus, DM consisting from PGB of familon
type contains a "hot" component from massless  particles and a "cold"
(nonrelativistic) component from massive particles.

\vspace*{.3cm}

\noindent
{\bf 4. THE THERMODYNAMICS OF A PHASE TRANSITION AND THE BLOCKLY-PHASE
STRUCTURE OF FAMILON GAS}

Since the preon vacuum is the nonlinear medium formed by the
strong chromodynamical and metachromodynamical interactions therefore
the strong nonlinearity of perturbations of this medium is evident. That
is the own nonlinearity of familon system provides firstly the existence
of RPT with spontaneous breaking of familon fields of residual symmetry
and secondly allows to describe RPT by the thermodynamical method.
Really, if the constants of bound $\lambda_{0}, \lambda_{1},
\lambda_{2}, \lambda_{12}$ in the Lagrangian (3.1) are not anomally
small then the familon gas created during evolution of the Universe in
the moment of production of metachromodynamical and preon condensates
must quckly relax to thermodynamical equilibrium state. Certainly, the
total thermodynamical equilibrium between familons and usual particles
(quarks, leptons, photons, gluons ...) can not be and the temperatures
of familon gas do not need to coincide with thermodynamical
temperatures of all other subsystems of the Universe. Thus our task is
to build the thermodynamics of the field system with Lagrangian (3.1)
by the method of the temperature quantum theory of field. In this task
the specific element takes place which is absent in early studying
models.

In the tree approximation spontaneous breaking of symmetry in Lagrangian
(3.1) is impossible (if this approxmation takes into attention then the
stable familon vacuum is absent). Also evidently that the
approximation is wrong for the task about RPT in familon gas because
of the scale of pseudogoldstone mass  more than ten orders smaller the
scale of metacolor confinement (for which this Lagrangian is deduced).
For lowering with up scale to  down scale the effect of renorm
group evolution follows to take into account. We have done the
theoretico-field investigation of Lagrangian (3.1) and have got the
result that stable vacuum arises only for realization of the strong
bound regime on the scale  of metacolour confinement. That is the strong
own nonlinearity of familon field is simultaneously the condition of
familon vacuum stability.

The pseudogoldstone part of terminally-renorm Lagrangian can be written
as:
$$L = \frac{1}{2} (\partial_{\mu} \varphi^{'} \partial^{\mu} \varphi^{'} -
m^{2}_{\varphi^{'}} \varphi^{'} \varphi^{'}) +  \partial_{\mu} \varphi^{+}
\partial^{\mu} \varphi - m^{2}_{\varphi} \varphi^{+}  \varphi + \partial_{\mu}
f^{+} \partial^{\mu} f - m^{2}_{f} f^{+} f -$$
$$- \frac{1}{2} \lambda_{1} (\varphi^{+} \varphi)^{2} - \frac{1}{2}
\lambda_{2} (f^{+} f)^{2} - \frac{1}{4} \lambda_{3} \varphi^{' 4} -
\lambda_{12} \varphi^{+} \varphi f^{+} f - \frac{1}{2} k_{12} (\varphi^{+}
f + f^{+} \varphi)^{2} - \eqno(4.1)$$
$$- \frac{1}{2} \varphi^{' 2} (\lambda_{13} \varphi^{+} \varphi + \lambda_{23}
f^{+} f) - \frac{1}{2} [M^{2}_{1} (\delta v)^{2} + M^{2}_{2} (\delta
u)^{2} + M^{2}_{12} \delta u \delta v ] -$$
$$- \delta v (h_{1} \varphi^{+} \varphi + h_{2} f^{+} f + \frac{1}{2}
h_{3} \varphi^{' 2}) - \delta u (k_{1} \varphi^{+} \varphi + k_{2} f^{+} f
+ \frac{1}{2} k_{3} \varphi^{' 2})$$
where $\varphi^{'}, \varphi = \frac{1}{\sqrt{2}} \; (\varphi_{1} + i \varphi_{2})$
are pseudoscalar familon fields, $f = \frac{1}{\sqrt{2}} \; (f_{1} + i f_{2})$
is the scalar familon field which drops out as a condensate for
superlow temperatures
$$\langle f \rangle = \frac{1}{\sqrt{2}}\; \langle f_{i} \rangle =
\frac{1}{\sqrt{2}} \; \eta \eqno(4.2)$$
(in the subsequent discussion $\eta$ is named as own parameter of order
PO), $\delta v$ and $\delta u$ are nonown PO which are responsible
for vacuum shifts of heavy familon fields induced by an own PO. All
constants in (4.1) are terminally-renorm version of other initial
constants of the Lagrangian or their combinations. (Some from constants
of bound have been marked by the letters $\lambda$ but they do not
coincide with analogous symbols in (3.1)).

The thermodynamics of the Lagrang system (3.1) was formulated in the
approximation of self-coordinated field [13]. The basic task is to find
Landau unequilibrium functional state that is the density of free
energy $F(\eta, T)$. For strong nonlinear systems, like to discussed,
the Landau functional failed to find in an explisit form. In a
nonexplisit form in the approximation of self-coordinated field the
Landau functional is given by a functional depending on the own and the
nonown PO and 5 effective masses of particles. Besides, to this
functional system nonlinear equations of bound are added allowing to
express the effective masses of particles through the parameter of
order and the temperature. Fortunately the calculation of all
observable values and finding of stability condition of phases can be
conducted in technics of a more simple functional which is in the form:
$$F(T, \eta, m_{11}, m_{12}, m_{3}, m_{21}, m_{22} = - \frac{1}{3}
\sum_{A} J_{2} (T, m_{A}) + U(\eta, m_{A}) \eqno(4.3)$$
$$\frac{1}{16 \pi^{2}} \; m^{4}_{11} ln \;
\frac{m^{2}_{11}}{m^{2}_{11(0)}} + J_{1} (T, m_{11}) + ... =$$
$$\frac{1}{16 \pi^{2}} \; m^{4}_{12} ln \;
\frac{m^{2}_{12}}{m^{2}_{12(0)}} + J_{1} (T, m_{12}) + ... =$$
$$\frac{1}{16 \pi^{2}} \; m^{4}_{3} ln \;
\frac{m^{2}_{3}}{m^{2}_{3(0)}} + J_{1} (T, m_{3}) + ... = \eqno(4.4)$$
$$\frac{1}{16 \pi^{2}} \; m^{4}_{21} ln \;
\frac{m^{2}_{21}}{m^{2}_{21(0)}} + J_{1} (T, m_{21}) + ... =$$
$$\frac{1}{16 \pi^{2}} \; m^{4}_{22} ln \;
\frac{m^{2}_{22}}{m^{2}_{22(0)}} + J_{1} (T, m_{22}) + ... =$$
Here $m_{11}$ and $m_{12}$ are the effective masses (depending on
temperature) of real pseudoscalar fields arisen in LS phase for decay
of the complex field $\phi$ (in HS phase $m_{11} = m_{12} = m_{1});
m_{12}$ and $m_{22}$ are analogous masses of real scalar fields arisen
in HS phase for decay of complex field $f$ (in HS phase $m_{21} =
m_{22} = m_{2}); m_{3}$ is the effective mass of the real pseudoscalar
field $\varphi^{'}$;
$$
J_{n} (T, m_{A}) = \frac{1}{2 \pi^2} \int \limits_p
\frac{p^{2n}dp}{\sqrt{p^2+ m^2_A}
( \exp  \frac{p^2+m^2_A}{T}  -1)}\;\; n = 1, 2, 3; A = 11,
12, 3, 21, 22
$$
is the characteristic integral through which expresses observable values.

The expession (4.3) can consider and as the performing functional when
equations of bound (4.4) lie on its extremalies in space of effective
masses. If this circumstance takes into attention then the necessary
condition of nonequilibrium functional minimum on own PO can write in
the form:
$$\eta (m^{2}_{21} - 2 \lambda_{2} \eta^{2}) = 0 \eqno(4.5)$$
which must be solved together with equations of bound (4.4). All
branches of solutions of this equations system are tistified on
equilibrium by the sufficient conditions of minimum:
$$
\frac{d^{2} F}{d \eta^{2}} = \frac{\partial^{2} F}{\partial \eta^{2}}
+ \sum_{A} \frac{\partial^{2} F}{\partial \eta \partial m_{A}} \;
(\frac{\partial m_A}{\partial \eta}) > 0 \eqno(4.6)
$$

The  system of equations (4.4) and (4.5) was being solved numerically
in the region temperatures $0 < T<
5 \mid m_{f} \mid$, where $\mid m_{f} \mid$ is defined by formula (3.9). The
solution for LS phase exists in the region of temperatures $0 < T < 2.5
\mid m_{f} \mid$ that is $T_{c(1)}$ is the upper thermodynamical
boundary of LS phase stability. This system of equations is integrated
independently for HS phase for $\eta = 0$. The HS phase exists for $T >
0.5 \mid m_{f} \mid$ that is $T_{c(2)} = 0.5 \mid m_{f} \mid$ is the
lower boundary of HS phase stability.

Thus the region of coexistence HS and LS phases is realized in wide
temperature interval $0.5 \mid m_{f} \mid < T < 2.5 \mid m_{f} \mid$.
By mathematically the coexistence of phases is expressed in presense of
two stable branches of solutions (4.4) and (4.5) for the same
thermodynamical parameters.  In the
middle of coexistence region for $T_{eq} = 1.5 \mid m_{f} \mid$ the
phase equilibrium point is fixed by equality of free energies $F(HS) =
F(LS)$. In the different phases the system possesses different
equations of state therefore the mechanical equilibrium between regions
of HS and LS phases which is reached for equality of their pressure can
be realized only for inequility of their density energies. On Fig. 1
the dependences of density energies $\epsilon^{HS} (p)$ and
$\epsilon^{LS} (p)$ in region of phase existence are shown. From Fig. 1
it can see the equilibrium phase in the region of their coexistence is
possible only for sharp contrast of the density
$$ \frac{2 \mid\epsilon^{(HS)} (p) - \epsilon^{(LS)} (p) \mid}
{\mid\epsilon^{(HS)} (p) + \epsilon^{(LS)} (p) \mid} > 1$$

Thus the cosmological familon gas evolves thermodynamically from the
region of HS phase stability in the region of LS phase stability. For
temperatures $T < T_{c(1)}$ the most part of familon gas which has been in
HS phase starts to arise seeds of LS phase. However these seeds are
unstable till to temperature of equilibrium of phases $T_{eq}$. In this
temperature region the spontaneous creation and death of seeds occur.
The production of seeds is the process of generation of density inhomogeneity
the development of which must come to gravitation instability of DM.
More strong the familon gas condenses in the temperature range
$T_{c(2)} < T < T_{eq}$. Here seeds of LS phase are stable and coexist
with rests of HS phase in regime of the density high contrast. Probably
that the more  dense HS phase in these conditions starts to collapse
gravitationally and to isolate from regions of the LS phase.

Thus at the region of RPT in familon gas catastrophic phenomena happen
one from consequenses which is temperary coexistense of blockly-phase
structure (the space alternation of HS and LS phase regions). The
characteristic scale of this structure is determined by the distance to
horizon events at the moment of RPT. The same scale certainly
determines and the characteristic scale of baryon subsystem structure which
reacts gravitationally on fragmentation of DM. If the heterogeneity of
familon gas takes into attention then this phenomena in the Universe
must be repeated as minimum two times for different temperatures of
familon gas that is for different sizes of horizon of events. Probably
this mechanism of DM  structurization allows better  to understand the
origin of scale hierarchy of the baryon component.

\vspace*{.3cm}

\noindent
{\bf 5. CONCLUSION}

Our model is unambiguously connected with preon model of elementary
particles having a good perspective of experimental check on colliders. The
experimental status for this model may get only after the detection of
familons in rare decays of heavy mesons. After getting of the experimental
status the reception of this model in cosmology will be inevitably. The
quality of this model with point of view of cosmology is explicit today.
The role of particle generations in the Universe become more evident. The
structurization of DM (and the baryon subsystem as consequence) gives
particles arising only for consideration of symmetry among generations.
That is for possibility of structurization of matter it is necessary as
minimum three generations of particles.

\vspace*{.3cm}

\noindent
{\bf References}

\noindent
1. I.M.~Hook, R.G.~McMahon.  MN, 294, L7 (1998)\\
2. A.~Omont et al. Nature 382, 428 (1996);
Guilloteau et al. Astron. Astrophys. 328, L1 (1997)\\
3. M.~Rees. Proc. Natl. Acad. Sci. USA, 95, 47 (1998): Ch.C.~Steidel.
Proc. Natl. Acad. Sci. USA 95, 22 (1998)\\
4. L.~Bergstrom. Nucl.Phys.B (Proc.Suppl.) 70,31(1999); D.O.~Caldwell.
Nucl. Phys. B (Proc.Suppl.) 70,43 (1999)\\
5. C.T.~Hill, D.N.Schramm, J.~Fry. Comments of Nucl. and Particle Phys.
19, 25 (1989)\\
6. J.A.~Frieman, C.T.~Hill, R.~Watkins. Preprint Fermilab Pub.-91/324-A (1991)\\
7. R.M.~Barnet et al. Phys. Rev. D54 (1996)\\
8. D.A.~Kirzhnits and A.D.~Linde. JETP 67, 1263 (1974)\\
9. M.Yu.~Khlopov, A.S.~Sakharov.  Phys.Atom.Nucl. 57,651 (1994)\\
10. H.~Fritzsch, G.~Mandelbaum. Phys. Lett. B 102, 319 (1981)\\
11. C.~Adloff et al. Z. Phys. C 74, 191 (1997); J.Z.~Breitweg. Phys. C
74, 207 (1997)\\
12. M.A.~Shifman, A.I.~Vainstein, V.I.~Zakharov. Nucl. Phys. B 147, 385 (1979)\\
13. G.~Vereshkov, V.~Burdyuzha. Intern. J. of Modern Phys. 10, 1343 (1995)\\

\end{document}